%% file: main.tex
\pdfoutput=1 
\documentclass[letterpaper]{article} 
\usepackage{aaai25}  
\usepackage{times}  
\usepackage{helvet}  
\usepackage{courier}  
\usepackage[hyphens]{url}  
\usepackage{graphicx} 
\usepackage{natbib}  
\usepackage{caption} 
\usepackage{booktabs}
\usepackage{xcolor}
\newcommand{\answerYes}[1]{\textcolor{blue}{#1}}
\newcommand{\answerNo}[1]{\textcolor{teal}{#1}}
\newcommand{\answerNA}[1]{\textcolor{gray}{#1}}

\nocopyright
\newcommand{\NIssues}{371}
\newcommand{\MaxIssue}{372}
\newcommand{\MissingIssue}{105}
\newcommand{\NVideos}{13{,}758}
\newcommand{\NUploaders}{4{,}407}
\newcommand{\NCategories}{115}
\newcommand{\NComments}{47{,}799{,}657}
\newcommand{\NCommentsShort}{47.8M}
\newcommand{\NCommenters}{13{,}978{,}794}

\newcommand{\NDanmaku}{125{,}690{,}760}
\newcommand{\NDanmakuShort}{125.7M}
\newcommand{\NDanmakuUsers}{23.9M}
\newcommand{\OffViews}{77.8 billion}
\newcommand{\OffReplies}{118{,}559{,}881}
\newcommand{\OffDanmaku}{327{,}133{,}309}
\newcommand{\DanmakuCoverage}{38.4\%}
\newcommand{\FirstIssue}{22 March 2019}
\newcommand{\LastIssue}{7 May 2026}
\newcommand{\CollectionStart}{12 May 2026}
\newcommand{\CollectionEnd}{11 August 2026}

\newcommand{\NCovers}{13{,}739}
\newcommand{\NCommentMismatch}{25}
\newcommand{\MedDuration}{385}
\newcommand{\HRECRef}{EA260679}

\input{release-info.tex}

\begin{document}

\title{BiliPicks: Seven Years of Weekly Platform Curation on Bilibili\\
with Video Metadata, Comments, and Danmaku}

\author{
    Ruomeng Liu
}
\affiliations{
    The University of Hong Kong\\
    ruomeng.6@connect.hku.hk
}

\maketitle

\begin{abstract}
BiliPicks documents seven years of Bilibili's \emph{Weekly Picks}, a platform-curated editorial video series. The dataset covers \NIssues{} publicly listed issues and \NVideos{} curated videos from \FirstIssue{} to \LastIssue{}. It links issue dates, themes and editorial recommendations to video metadata, \NComments{} top-level comments, \NDanmaku{} time-synchronised overlay comments (danmaku), and \NCovers{} cover images. Weekly issues provide a consistent unit for tracing changes in category composition, recurring creators and editorial framing. Corresponding comments and danmaku connect video-level discussion with audience expressions tied to particular moments, while covers add a visual dimension. We evaluate retrieval coverage and illustrate longitudinal analyses of selection, audience language and cover presentation. BiliPicks provides a dataset for future studies of platform curation, linking changes in published selections to their editorial framing and corresponding audience engagement.
\end{abstract}

\noindent\textbf{Dataset:} \expandafter\url\expandafter{\DatasetURL}\par

\section{Introduction}

Platforms curate visibility through recommendations, rankings and editorial collections. These choices shape which creators and cultural forms receive attention \citep{gillespie2014relevance,bonini2019first}. Studying curation over time requires dated records of what a platform selected and how it presented those selections. A recurring editorial series provides such a record, with each issue identifying a set of promoted videos and the language used to recommend them.

Bilibili's \emph{Weekly Picks} (\emph{Meizhou Bikan}) publishes numbered issues with dates, themes and short recommendations for individual videos \citep{bilibiliWeeklyIndex,bilibiliWeeklyIssue1}. We introduce BiliPicks, a dataset linking this editorial record to the corresponding video metadata, audience messages and cover images. It covers \NIssues{} publicly listed issues from 2019 to 2026.\footnote{Bilibili's public index omits issue~\MissingIssue{}, yielding \NIssues{} listed issues across \MaxIssue{} issue numbers.}

BiliPicks contributes a longitudinal record of platform-designed curation at weekly resolution. Researchers can compare the genres and creators featured across successive issues, examine changes in recommendation language, and connect those changes to visual presentation. The same videos carry two forms of audience engagement: asynchronous comments and danmaku tied to playback positions. Their correspondence supports comparisons of how audiences discuss a video as a whole and respond to particular moments within it.

The dataset combines these dimensions within a single dated sampling frame. Existing video datasets commonly sample through queries, popularity rankings or creator lists; Table~\ref{tab:related} situates BiliPicks among Chinese video-engagement datasets. We evaluate collection coverage and illustrate research uses through changes in category composition, audience language and cover appearance. The linked selections, messages and images support future studies of editorial selection and its changing cultural context.

\begin{table*}[t]
\centering
\caption{
Representative corpora of audience engagement on Bilibili and
other Chinese video platforms. Text-unit counts are rounded where
appropriate.
}
\label{tab:related}

\small
\setlength{\tabcolsep}{2pt}
\renewcommand{\arraystretch}{1.12}

\begin{tabular}{
    @{}
    p{3.8cm}
    l
    r
    r
    p{5.5cm}
    l
    @{}
}
\toprule
\raggedright Dataset
    & Platform
    & Videos
    & Text units
    & \raggedright Sampling frame / span
    & Engagement
    \tabularnewline
\midrule

\raggedright TSC \citep{tsc2017}
    & Bilibili
    & 7{,}166
    & 11.84M
    & \raggedright Movie category; through 2015
    & Danmaku
    \tabularnewline

\raggedright VideoIC \citep{videoic2020}
    & Bilibili
    & 4{,}951
    & 5.33M
    & \raggedright Popular-category sample
    & Danmaku
    \tabularnewline

\raggedright DanmakuTPPBench \citep{danmakutpp2025}
    & Bilibili
    & 7{,}250
    & 10.82M
    & \raggedright Creators in the 2024 top-100 list
    & Danmaku
    \tabularnewline

\raggedright LCAffect \citep{deng2025enhancing}
    & Multi-platform
    & 2{,}966
    & 11.36M
    & \raggedright User-generated content, television, and films
    & Timed comments
    \tabularnewline

\raggedright Graphicon \citep{zhang2022graphicon}
    & Bilibili
    & 42
    & 1.03M
    & \raggedright Spring Festival Gala channel, 2010--2022
    & Async.\ comments
    \tabularnewline

\raggedright HOTVCOM \citep{chen2024hotvcom}
    & Douyin
    & 94k
    & 137M
    & \raggedright Short-video sample
    & Async.\ comments
    \tabularnewline

\midrule

\raggedright \textbf{BiliPicks (ours)}
    & \textbf{Bilibili}
    & \textbf{\NVideos}
    & \textbf{\shortstack[r]{\NCommentsShort{}\\
                              + \NDanmakuShort}}
    & \raggedright\textbf{\emph{Weekly Picks}, 2019--2026}
    & \textbf{Danmaku + async.}
    \tabularnewline

\bottomrule
\end{tabular}

\vspace{2pt}

\begin{minipage}{0.98\textwidth}
\footnotesize
\raggedright
\emph{Note.}
TSC statistics follow the full corpus reported in the original paper.
\end{minipage}
\end{table*}

\section{Related Work}

\subsection{Platform Curation over Time}

Curation involves selecting content, arranging it for attention and supplying a frame through which audiences encounter it. In news, \citet{bruns2022gatewatching} examines how professional and audience curation operate within social media, while \citet{diakopoulos2015editor} studies editorial judgements of comment relevance. \citet{gillespie2014relevance} connects algorithmic relevance to the institutional choices behind information selection. These perspectives direct attention to both the items a platform promotes and the ways it presents them.

Editorial and algorithmic curation can also work together. \citet{bonini2019first} describe how music-streaming gatekeepers combine editorial judgement, audience metrics and algorithmic distribution. On Apple News, \citet{bandy2020auditing} compare human-edited \emph{Top Stories} with algorithmic \emph{Trending Stories}, finding differences in source concentration, diversity and soft-news prevalence. For video platforms, \citet{goncalves2026youtube} archive YouTube Trending across 104 countries over three years. Together, these studies motivate measuring the published outputs of curation across settings and time.

Bilibili provides a setting in which editorial promotion, creator participation and platform governance intersect. \citet{wang2022popularising} documents how campaigns, monetisation and editorial recommendations shaped the growth of vlogging. Work on community governance examines Bilibili's mediation between users and the state \citep{chen2023governing} and its participatory content-moderation system \citep{shang2026shifting}. \citet{lu2026recommending} examine algorithmic promotion of state-created content through trending and recommendation systems. This research shows why Bilibili's published selections warrant study alongside its recommendation algorithms.

\emph{Weekly Picks} provides an issue-based editorial record for that purpose. Each issue groups videos under a theme and attaches a recommendation to each entry. Its weekly resolution and seven-year span support comparisons of which genres and creators recur, how recommendation language changes, and how those patterns relate to audience expression. Following the same editorial series across successive issues connects institutional curation to observable changes in featured content and framing.

\subsection{Video and Audience-Engagement Datasets}
\label{sec:related_datasets}

Sampling choices determine what a video dataset can describe. \citet{ribeiro2021youniverse} review YouTube sampling through search, recommendation-based snowballing and creator lists. On Bilibili, LiveBot uses category-spanning queries \citep{ma2019livebot}, \citet{he2021beyond} study popular crisis videos, and DanmakuTPPBench samples prominent creators \citep{danmakutpp2025}.

Table~\ref{tab:related} compares resources containing audience-authored text. TSC, VideoIC, DanmakuTPPBench and LCAffect primarily contain video-timed messages \citep{tsc2017,videoic2020,danmakutpp2025,deng2025enhancing}; a smaller Bilibili dataset supplies sentiment annotations \citep{danmakusentiment2024}. Graphicon follows asynchronous comments and replies on 42 videos over 2010--2022 \citep{zhang2022graphicon}, while HOTVCOM collects asynchronous comments on Douyin at larger scale \citep{chen2024hotvcom}. Livestream chats capture another temporal setting \citep{jiang2024games}. BiliPicks adds corresponding comments and danmaku within a recurring editorial selection, linking audience expression to issue-level change.


\begin{figure*}[t]
\centering
\includegraphics[width=0.92\textwidth]{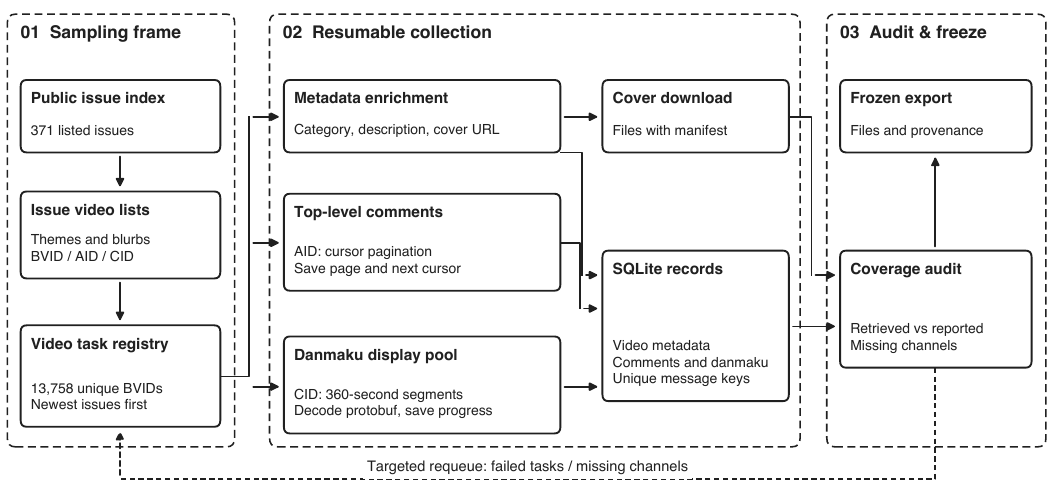}
\caption{Collection of linked issue, video, audience and cover data. The three dashed frames group the sampling frame, resumable collection and audit of the source snapshot. Solid arrows denote task and data dependencies; dashed arrows denote targeted requeueing.}
\label{fig:dataset}
\end{figure*}

\begin{figure*}[t]
\centering
\includegraphics[width=\textwidth]{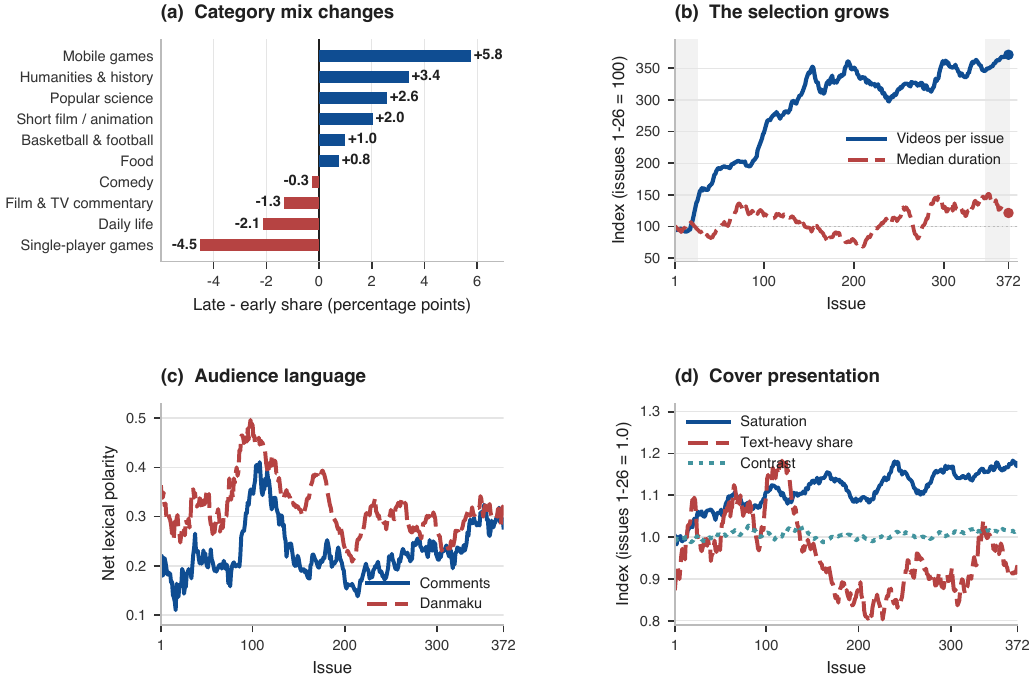}
\caption{Selection and presentation across weekly issues. (a) Category-share changes between the first and last 26 issues. (b) Issue size and video duration, indexed to issues 1--26. (c) Net lexical polarity in sampled comments and danmaku. (d) Saturation, contrast and text-heavy cover share, indexed to issues 1--26. Sampling, smoothing and measurement definitions are in Appendix~\ref{sec:measurement_details}.}
\label{fig:application}
\end{figure*}

\begin{figure*}[t]
\centering
\includegraphics[width=\textwidth]{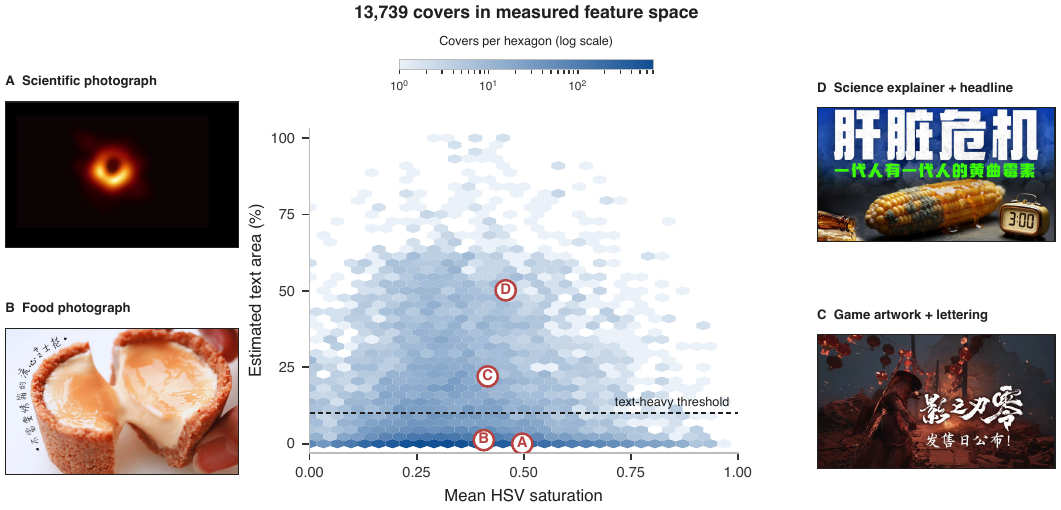}
\caption{Cover appearance across \NCovers{} images. Hexagons show counts on a logarithmic scale. Four source images, shown at their original aspect ratios, illustrate positions in the saturation and estimated text-area distribution. The dashed line marks the text-heavy threshold of 0.10. Measurement details are in Appendix~\ref{sec:measurement_details}.}
\label{fig:cover_cases}
\end{figure*}

\section{Dataset Construction}

\paragraph{Sampling frame.}
We collected the publicly listed \emph{Weekly Picks} archive from \CollectionStart{} to \CollectionEnd{}. The platform's issue index defined the sampling frame: we enumerated every listed issue, retrieved its video membership and editorial recommendations, and built a video task registry. Figure~\ref{fig:dataset} shows how this registry connects collection of metadata, audience messages and cover images. Collection used a mainland-China request origin.

\paragraph{Video metadata and covers.}
For each video, we retrieved uploader attribution, categories, duration, publication time and engagement counters. These counters describe activity observed during collection. The published cover URLs yielded \NCovers{} images; the remaining 19 videos had no cover URL recorded. Video identifiers link these assets to their weekly issues and audience messages.

\paragraph{Audience messages.}
We retrieved top-level comments through cursor-based pagination, applying the same scope across videos. Nested replies are excluded, whereas the platform's official reply counter includes them. Danmaku were retrieved from the display pool in 360-second playback segments, preserving their positions within each video. The resulting channels connect discussion of a video as a whole with expressions tied to particular moments.

\paragraph{Recovery and coverage.}
Persistent checkpoints recorded video tasks, comment cursors and attempted danmaku segments. Source-message keys prevented duplicates during retries, while failed tasks and videos with reported activity but no retrieved messages could be requeued. Shared rate limiting and backoff controlled request frequency. We assessed coverage by comparing retrieved records with the platform's reported counts; the results are reported in Data Quality and Interpretation. Appendix~\ref{sec:collection_details} gives endpoints, request settings and checkpoint behaviour.

\section{Dataset Description}

BiliPicks links issue membership and editorial recommendations to video metadata, two audience-message channels and cover images. Table~\ref{tab:schema} describes these layers, and Table~\ref{tab:stats} summarises their scale. Issue and video identifiers connect the files, allowing analyses to move between individual messages, videos and weekly selections.

\begin{table}[t]
\centering
\caption{BiliPicks structure. Video and issue identifiers link the layers; the README explains joins and selected fields.}
\label{tab:schema}
\small
\begin{tabular}{lp{5.15cm}}
\toprule
Layer & Contents \\
\midrule
Issues & Issue number, label, theme and dates \\
Videos & Video and issue IDs; title, uploader attribution, category, publication time, duration, recommendation text, counters and collection time \\
Comments & Release row ID, video ID, posting date, likes, text and text-processing status \\
Danmaku & Release row ID, video/page IDs, video position, posting date, display fields, text and text-processing status \\
Covers & Video ID, image filename, archive, dimensions, checksum and review status \\
Derived data & Per-video counts, sampled lexical features, issue-level summaries and visual measurements \\
\bottomrule
\end{tabular}
\end{table}

\begin{table}[!b]
\centering
\caption{BiliPicks v1.0: collection statistics. Account and author-hash counts describe source records, before identifier removal.}
\label{tab:stats}
\small
\begin{tabular}{lr}
\toprule
Issues (all publicly listed) & \NIssues \\
Videos & \NVideos \\
Uploaders & \NUploaders \\
Observed platform category labels & \NCategories \\
Videos per issue (min / mean / max) & 8 / 37.1 / 50 \\
Median video duration & \MedDuration\,s \\
First issue & \FirstIssue \\
Last issue & \LastIssue \\
Collection began & \CollectionStart \\
Collection ended & \CollectionEnd \\
Cover images & \NCovers \\
\midrule
Top-level comments & \NComments \\
Distinct source commenter accounts & \NCommenters \\
Danmaku & \NDanmaku \\
Distinct source danmaku author hashes & \NDanmakuUsers \\
Videos with $\geq$1 comment & 13{,}725 \\
Videos with $\geq$1 danmaku & 13{,}735 \\
\midrule
Official views (sum) & \OffViews \\
Official replies (sum) & \OffReplies \\
Official danmaku (sum) & \OffDanmaku \\
\bottomrule
\end{tabular}
\end{table}

The weekly series changes in both size and composition. Between the first and last half-year, mean videos per issue increase from 13.0 to 46.8, and mean per-issue median duration rises from 365 to 497 seconds (Figure~\ref{fig:application}b). Mobile games, humanities and history, and popular science gain the most category share, while single-player games decline most (Figure~\ref{fig:application}a). These comparisons illustrate how weekly issue membership makes changes in editorial selection measurable.


\section{Data Quality and Interpretation}

\paragraph{Retrieval coverage.}
We retrieved \NDanmakuShort{} of \OffDanmaku{} officially counted danmaku, an overall coverage of \DanmakuCoverage{}. Coverage depends on video popularity: median retrieval is 98.2\% below 1,000 officially counted messages and 25.6\% above 50,000. The display pool therefore represents heavily discussed videos less completely. \NCommentMismatch{} videos have non-zero official reply counts but no retrieved top-level comments; fourteen have non-zero danmaku counts but no retrieved danmaku. Appendix~\ref{sec:coverage_details} gives the full breakdown. Top-level comments amount to about 40\% of the official reply counter, which also includes nested replies.

\paragraph{Timing and sampling frame.}
Weekly issue dates provide the longitudinal structure; engagement counters are snapshots from collection in 2026. Older videos have had more time to accumulate activity. Comment timestamps span 17 March 2019 to 10 August 2026, including discussion after selection. Publication and collection times support age-aware comparisons, although the final record-update time can be later than the counter observation. BiliPicks describes the platform's published editorial series; estimating the effect of being featured requires a separate causal design.

\paragraph{Concentration.}
A median video carries 2,494 comments and 5,636 danmaku. One video in issue 357 contributes 4.52M comments, or 9.5\% of all collected comments. Video-level comparisons should therefore consider the influence of highly discussed videos, while issue-level analyses should account for changes in selection size.

\section{Research Applications}

BiliPicks supports research on how platform curation and audience expression change together. The following analyses illustrate questions enabled by its linked structure; measurement and sampling details are in Appendix~\ref{sec:measurement_details}.

\paragraph{Editorial selection.}
Issue membership supports comparisons of category diversity, recurring creators and recommendation language. The expansion and category shifts in Figure~\ref{fig:application}(a--b) show that the editorial series itself changes over time. Researchers can examine whether its recommendation language changes alongside the genres and creators it features, or compare its selections with independently collected trending lists.

\paragraph{Audience language.}
We compare corresponding comment and danmaku samples using HowNet lexical polarity through \texttt{cnsenti} \citep{deng2020cnsenti}. In 111,820 comments and 147,945 danmaku, polarity terms occur in 49.8\% and 27.4\% of messages, respectively. Mean issue-level polarity rises from 0.154 to 0.282 for comments and from 0.284 to 0.310 for danmaku between the first and last 26 issues (Figure~\ref{fig:application}c). Comparisons track each channel's trajectory because lexical coverage differs.

\paragraph{Visual presentation.}
Cover measurements connect changes in visual presentation to the same weekly selections. In the final 26 issues, mean saturation is 16.7\% higher than in issues 1--26, heavily captioned share is 6.9\% lower, and contrast is 1.6\% higher (Figure~\ref{fig:application}d). Figure~\ref{fig:cover_cases} places four source-image examples within the full distribution of saturation and estimated text area. Linking covers, editorial recommendations and corresponding messages supports studies of how visual and textual presentation relate to audience expression.

\section{Data Availability and Ethical Considerations}

\paragraph{Availability.}
\DatasetAvailability
The README explains the dataset's structure, processing and reuse conditions, following datasheet and FAIR documentation principles \citep{gebru2021datasheets,wilkinson2016fair}. Code and derived tables support reproduction of the reported figures.

\paragraph{Ethics and privacy.}
HKU's Human Research Ethics Committee approved the project (\HRECRef{}). We collected public material without recruitment, contact or intervention; no individual consent was obtained. Processing removes audience account identifiers and source message IDs, reduces posting times to UTC dates, and withholds flagged text while retaining message records. The processed dataset retains non-empty text for \NReleasedCommentTexts{} comments and \NReleasedDanmakuTexts{} danmaku; \NWithheldTexts{} texts are withheld and \NEmptyTexts{} source rows were empty. Image metadata is stripped while preserving pixels. Public uploader attribution remains, and searchable quotations and images can identify people. Reuse terms prohibit re-identification, contacting and profiling; the README states third-party rights and a correction/removal contact. Appendix~\ref{sec:processing_details} documents the processing rules.

\section{Conclusion}

BiliPicks provides a dataset for longitudinal studies of platform curation at weekly resolution. Seven years of issue membership and editorial recommendations connect the platform's published choices to corresponding metadata, audience messages and cover images. This structure supports research on changes in featured genres and creators, editorial framing, audience language and visual presentation.

\section{Acknowledgement}

The author thanks Xizhu Zeng for helpful comments on data curation. The author used ChatGPT (OpenAI) and Claude (Anthropic) to assist with research, table and citation formatting, and manuscript readability. The author reviewed and approved all final content.

\bibliography{refs}

\clearpage
\section{Paper Checklist}
\begingroup
\small
\begin{enumerate}
\setlength{\itemsep}{0pt}
\setlength{\parsep}{0pt}

\item For most authors...
  \begin{enumerate}
  \setlength{\itemsep}{0pt}
  \item Would answering this research question advance science without
  violating social contracts, such as violating privacy norms, perpetuating
  unfair profiling, exacerbating the socio-economic divide, or implying
  disrespect to societies or cultures? \answerYes{Yes. The paper studies a public
  promotional surface and uses population-level reporting, data minimisation, and an
  approved open-release scope to reduce privacy and redistribution risks.}
  \item Do your main claims in the abstract and introduction accurately reflect
  the paper's contributions and scope? \answerYes{Yes.}
  \item Do you clarify how the proposed methodological approach is appropriate
  for the claims made? \answerYes{Yes. See Dataset Construction, Data Quality and Interpretation, and Research Applications.}
  \item Do you clarify what are possible artifacts in the data used, given
  population-specific distributions? \answerYes{Yes. The paper discusses the weekly
  selection frame, collection origin, snapshot counters and popularity-dependent retrieval.}
  \item Did you describe the limitations of your work? \answerYes{Yes. See Data Quality and Interpretation.}
  \item Did you discuss any potential negative societal impacts of your work?
  \answerYes{Yes. See Data Availability and Ethical Considerations.}
  \item Did you discuss any potential misuse of your work? \answerYes{Yes. The
  paper addresses re-identification, contacting and profiling, and third-party reuse.}
  \item Did you describe steps taken to prevent or mitigate potential negative
  outcomes of the research, such as data and model documentation, data
  anonymization, responsible release, access control, and the reproducibility
  of findings? \answerYes{Yes. The paper describes data minimisation, removal of audience identifiers, text screening, date generalisation and reproducible figure code.}
  \item Have you read the ethics review guidelines and ensured that your paper
  conforms to them? \answerYes{Yes. The project received HKU HREC approval
  (reference \HRECRef{}), and the paper follows the approved protocol and
  release scope.}
  \end{enumerate}

\item Additionally, if your study involves hypotheses testing...
  \begin{enumerate}
  \setlength{\itemsep}{0pt}
  \item Did you clearly state the assumptions underlying all theoretical
  results? \answerNA{NA. The paper reports descriptive analyses and does not test
  theoretical hypotheses.}
  \item Have you provided justifications for all theoretical results?
  \answerNA{NA.}
  \item Did you discuss competing hypotheses or theories that might challenge
  or complement your theoretical results? \answerNA{NA.}
  \item Have you considered alternative mechanisms or explanations that might
  account for the same outcomes observed in your study? \answerNA{NA.}
  \item Did you address potential biases or limitations in your theoretical
  framework? \answerNA{NA.}
  \item Have you related your theoretical results to the existing literature in
  social science? \answerNA{NA.}
  \item Did you discuss the implications of your theoretical results for
  policy, practice, or further research in the social science domain?
  \answerNA{NA.}
  \end{enumerate}

\item Additionally, if you are including theoretical proofs...
  \begin{enumerate}
  \setlength{\itemsep}{0pt}
  \item Did you state the full set of assumptions of all theoretical results?
  \answerNA{NA. The paper contains no theoretical proofs.}
  \item Did you include complete proofs of all theoretical results? \answerNA{NA.}
  \end{enumerate}

\item Additionally, if you ran machine learning experiments...
  \begin{enumerate}
  \setlength{\itemsep}{0pt}
  \item Did you include the code, data, and instructions needed to reproduce
  the main experimental results (either in the supplemental material or as a
  URL)? \answerNA{NA. The paper does not train or evaluate a machine-learning
  model; it reports descriptive metadata, lexicon scores and low-level image measurements.}
  \item Did you specify all the training details (e.g., data splits,
  hyperparameters, how they were chosen)? \answerNA{NA.}
  \item Did you report error bars (e.g., with respect to the random seed after
  running experiments multiple times)? \answerNA{NA.}
  \item Did you include the total amount of compute and the type of resources
  used (e.g., type of GPUs, internal cluster, or cloud provider)? \answerNA{NA.}
  \item Do you justify how the proposed evaluation is sufficient and
  appropriate to the claims made? \answerNA{NA.}
  \item Do you discuss what is ``the cost'' of misclassification and fault
  (in)tolerance? \answerNA{NA.}
  \end{enumerate}

\item Additionally, if you are using existing assets (e.g., code, data,
models) or curating/releasing new assets, without compromising anonymity...
  \begin{enumerate}
  \setlength{\itemsep}{0pt}
  \item If your work uses existing assets, did you cite the creators?
  \answerYes{Yes. The paper cites the affect resource and all comparison
  datasets used in its claims.}
  \item Did you mention the license of the assets? \answerNo{No. The HowNet and
  DUTIR lexicons are not redistributed; their full licensing terms are not
  provided here.}
  \item Did you include any new assets in the supplemental material or as a
  URL? \DatasetAssetAnswer
  \item Did you discuss whether and how consent was obtained from people whose
  data you're using/curating? \answerYes{Yes. No individual consent was obtained;
  the paper explains the public collection context, ethics approval, and
  release safeguards.}
  \item Did you discuss whether the data you are using/curating contains
  personally identifiable information or offensive content? \answerYes{Yes. See Ethics and privacy.}
  \item If you are curating or releasing new datasets, did you discuss how you
  intend to make your datasets FAIR (see FORCE11 (2020))? \answerYes{Yes. See Availability and the accompanying documentation.}
  \item If you are curating or releasing new datasets, did you create a
  Datasheet for the Dataset (see Gebru et al. (2021))? \answerYes{Yes. The README incorporates a concise datasheet covering composition,
  sampling, processing, access, reuse and maintenance; this paper supplies
  the collection methods and ethical considerations.}
  \end{enumerate}

\enlargethispage{4\baselineskip}
\item Additionally, if you used crowdsourcing or conducted research with human
subjects, without compromising anonymity...
  \begin{enumerate}
  \setlength{\itemsep}{0pt}
  \item Did you include the full text of instructions given to participants and
  screenshots? \answerNA{NA. No participants were recruited and no instructions
  were issued.}
  \item Did you describe any potential participant risks, with mentions of
  Institutional Review Board (IRB) approvals? \answerYes{Yes. The paper describes
  risks to message authors and reports HKU HREC approval (\HRECRef{}).}
  \item Did you include the estimated hourly wage paid to participants and the
  total amount spent on participant compensation? \answerNA{NA. No participants
  were recruited or compensated.}
  \item Did you discuss how data is stored, shared, and deidentified?
  \answerYes{Yes. See Availability, Ethics and privacy, and Appendix~\ref{sec:processing_details}.}
  \end{enumerate}

\end{enumerate}
\endgroup

\clearpage
\appendix
\setcounter{secnumdepth}{2}
\section{Implementation and Measurement Details}
\label{sec:collection_details}

\subsection{Collection Sources and Recovery}

Under \url{https://api.bilibili.com}, \path{/x/web-interface/popular/series/list} supplies the issue index and \path{/x/web-interface/popular/series/one?number=N} supplies each issue's entries.

Collection used one worker on a mainland-China virtual machine, targeting 0.5 requests per second over 91 days. Requests shared rate limiting, exponential backoff and cooldown after platform risk responses. Comments were retrieved page by page using next-offset cursors. Nested replies were excluded consistently across videos.

Danmaku were decoded from protobuf responses in 360-second playback segments. SQLite checkpoints supported recovery, source-message keys prevented duplicates, and videos with reported activity but no retrieved messages could be requeued. Failed danmaku segments advanced the attempted-segment checkpoint, allowing gaps even after traversal completed. Coverage is therefore evaluated from retrieved counts.

\subsection{Coverage and Dataset Profile}
\label{sec:coverage_details}

Median danmaku retrieval is 98.2\% below 1,000 officially counted messages, 82.5\% for 1,000--10,000, 62.2\% for 10,000--50,000, and 25.6\% above 50,000. Besides the \NCommentMismatch{} videos with non-zero reply counts but no comments, eight have zero official replies and no retrieved comments. Fourteen have non-zero danmaku counts but no retrieved danmaku; individual causes were not verified.

The leading categories are comedy (1,296 videos), single-player games (887), daily life (784), mobile games (718), and short films and hand-drawn animation (565). Account/hash counts in Table~\ref{tab:stats} describe source records before identifier removal.

\newpage
\subsection{Sampling and Measurements}
\label{sec:measurement_details}

The systematic message sample selected nonempty messages by source IDs modulo 500 for comments and 1,300 for danmaku, then retained every second exported row. Source-ID gaps preclude treating this as simple random sampling. Analyses use the source sample before disclosure screening.

HowNet counts through \texttt{cnsenti} define net polarity as $(\mathrm{positive}-\mathrm{negative})/(\mathrm{positive}+\mathrm{negative})$, missing when the denominator is zero. Figure~\ref{fig:application} uses rolling smoothers of 13 issues for selection size/duration, 21 for polarity and 27 for covers. Size, duration and cover series are indexed to their means in issues 1--26.

OpenCV features use a 512-pixel working width. Brightness/contrast are L* mean/standard deviation; saturation is mean HSV saturation; text area estimates the fraction occupied by morphological text-like blocks. Heavily captioned share uses text area above 0.10. Case provenance records source URLs, uploader attribution, coordinates and image hashes.

\subsection{Processing and Reuse}
\label{sec:processing_details}

Disclosure screening flags contact details, account mentions and numeric identifiers; flagged text is blanked and its status recorded. Numerical derivatives retain their original sample definitions. The HowNet/DUTIR lexicons are not redistributed. Detailed provenance and checksums remain with the depositor.

\end{document}

%% file: release-info.tex
\newcommand{\DatasetURL}{https://doi.org/10.7910/DVN/RCDO59}
\newcommand{\NReleasedCommentTexts}{45{,}846{,}264}
\newcommand{\NReleasedDanmakuTexts}{125{,}224{,}504}
\newcommand{\NWithheldTexts}{2{,}419{,}594}
\newcommand{\NEmptyTexts}{55}
\newcommand{\DatasetAvailability}{%
  The Harvard Dataverse deposit is under repository review; its DOI is given
  below the abstract.\space}
\newcommand{\DatasetAssetAnswer}{%
  \answerNo{No. The dataset DOI and repository URL are provided, but public
  file access is pending repository publication.}}